# Fourth-Order Topological Insulator via Dimensional Reduction

Kai Chen[1,2], Matthew Weiner[1,2], Mengyao Li[1,2], Xiang Ni[3], Andrea Alù[3,2,1], and Alexander B. Khanikaev[1,2]

[1]Department of Electrical Engineering, Grove School of Engineering, City College of the City University of New York, 140th Street and Convent Avenue, New York, NY 10031, USA.

[2]Physics Program, Graduate Center of the City University of New York, New York, NY 10016, USA.

[3]Photonics Initiative, Advanced Science Research Center, City University of New York, New York, NY 10031, USA

**Abstract**

The properties of topological systems are inherently tied to their dimensionality. Higher-dimensional physical systems exhibit topological properties not shared by their lower dimensional counterparts and, in general, offer richer physics. One example is a *d*-dimensional quantized multipole topological insulator, which supports multipoles of order up to $2^d$ and a hierarchy of gapped boundary modes with 0-D corner modes at the top. While multipole topological insulators have been successfully realized in electromagnetic and mechanical 2D systems with quadrupole polarization, and a 3D octupole topological insulator was recently demonstrated in acoustics, going beyond the three physical dimensions of space is an intriguing and challenging task. In this work, we apply dimensional reduction to map a 4D higher-order topological insulator (HOTI) onto an equivalent aperiodic 1D array sharing the same spectrum, and emulate in this system the properties of a hexadecapole topological insulator. We observe the 1D counterpart of zero-energy states localized at 4D HOTI corners – the hallmark of multipole topological phase. Interestingly, the dimensional reduction guarantees that one of the 4D corner states remains localized to the edge of the 1D array, while all other localize in the bulk and retain their zero-energy eigenvalues. This discovery opens new directions in multi-dimensional topological physics arising in lower-dimensional aperiodic systems, and it unveils highly unusual resonances protected by topological properties inherited from higher dimensions.

**Introduction**

The dimensionality of a topological system plays a determining role in defining the symmetry classification and, therefore, the topological invariants that identify the topological phase specific to such system[1, 2, 3]. For instance, four-dimensional quantum Hall systems can exhibit a non-vanishing 2$^{nd}$ class Chern number that is not shared by systems with three or fewer dimensions, hence it cannot be implemented in three physical dimensions without mapping to a lower dimensional analogue[4]. However, already three-dimensional topological materials imply challenging fabrication demands due to the complex structures of the lattices[5, 6, 7, 8, 9]. It would therefore be highly advantageous to explore higher-dimensional topological physics with fewer physical spatial dimensions. This quest can be achieved in two ways: i) introducing synthetic dimensions or ii) mapping a higher-dimensional system onto its lower-dimensional counterpart. The first approach implements lattices with dimensions higher than the spatial dimensions by exploiting internal degrees of freedom, which could be of spectral[10, 11, 12, 13, 14, 15, 16, 17], temporal or spatial in nature[2, 3, 4, 5,18, 19, 20, 21], and therefore it remains experimentally challenging. The second approach, based on dimensional reduction and mapping onto a lower-dimensional system, has been proven quite fruitful in other areas of research: as an example, the celebrated Harper-Hofstadter Hamiltonian has been recently emulated in reconfigurable quasi-periodic 1D resonant acoustic lattices[22, 23]. In addition, the existence of boundary modes stemming from the 2nd Chern class topological phase has been reported in photonics[4] and in an angled optical superlattice of ultracold bosonic atoms[24].

Higher dimensions are crucial to unveil the new physics in the recently introduced class of higher-order topological insulators[25, 26]. The dimensionality of HOTIs defines the multiplicity of hosted topological boundary modes, which can be tailored on demand and protected by higher-dimensional symmetries. Moreover, the possibility to control both localization and spectral position of a large number of topologically protected boundary states can be truly transformative for many applications. So far the research on HOTIs mainly focused on the experimental realization of a 2D quadrupole topological phase[27, 28, 29, 30, 31, 32, 33]. The experimental realization of a 3D octupole topological phase remains challenging, due to the required complex dimerized coupling protocol within the lattice and the requirement of a synthetic magnetic flux of $\pi$ in all three dimensions. Only recently an octupole HOTI was implemented in 3D acoustic metamaterials fabricated by highly precise and controllable stereolithographic 3D printing[34], but translating these in a different physical domain and for shorter wavelengths would be extremely challenging.

In this work, we apply dimensionality reduction to demonstrate the crucial role of higher-dimensional topological physics by implementing a 4D hexadecapolar HOTI (h-HOTI), which is mapped onto its aperiodic 1D counterpart, and we experimentally confirm localization of projected topological 4D corner states in one dimension. To this aim, we apply Lanczos tridiagonalization and map a 4D tight-binding model of h-HOTI onto a one-dimensional acoustic system with strictly local aperiodic coupling distributions along the array. We obtain one pinned corner state of the 4D HOTI being preserved onto the edge of our sample by such mapping – a unique feature enabled by the Lanczos transformation[35, 36, 37, 38]. Also, the other corner states, despite being located in the bulk, are pinned in frequency and location, and preserve their topological characteristics, being protected by higher-dimensional symmetries.

## Results

The proposed 4D h-HOTI is described by a tight-binding Hamiltonian with nearest-neighbor coupling in 4D hypercubic unit cell, as schematically shown in Fig. 1**a**. In Bloch representation, the Hamiltonian can be expressed as

$$\widehat{\mathcal{H}}_{4D}(\boldsymbol{k}) = t_2 \sin(k_x) \hat{\Gamma}_3 + [t_1 + t_2 \cos(k_x)]\hat{\Gamma}_4 + t_2 \sin(k_y) \hat{\Gamma}_1$$
$$+[t_1 + t_2 \cos(k_y)]\hat{\Gamma}_2 + t_2 \sin(k_z) \hat{\Gamma}_5 + [t_1 + t_2 \cos(k_z)]\hat{\Gamma}_6$$
$$+t_2 \sin(k_w) \hat{\Gamma}_7 + [t_1 + t_2\cos(k_w)]\hat{\Gamma}_8, \qquad (1)$$

where $t_1$ and $t_2$ are the nearest-neighbor intracell and intercell hopping amplitudes, respectively. In Eq (1), the matrices $\hat{\Gamma}_i = -\sigma_3 \otimes \sigma_3 \otimes \sigma_2 \otimes \sigma_i$ for i = 1,2,3 , $\hat{\Gamma}_4 = \sigma_3 \otimes \sigma_3 \otimes \sigma_1 \otimes \sigma_0$, $\hat{\Gamma}_5 = \sigma_3 \otimes \sigma_2 \otimes \sigma_0 \otimes \sigma_0$ , $\hat{\Gamma}_6 = \sigma_3 \otimes \sigma_1 \otimes \sigma_0 \otimes \sigma_0$ , $\hat{\Gamma}_7 = \sigma_1 \otimes \sigma_0 \otimes \sigma_0 \otimes \sigma_0$ , $\hat{\Gamma}_8 = \sigma_2 \otimes \sigma_0 \otimes \sigma_0 \otimes \sigma_0$, $\sigma_{i=1,2,3}$ are Pauli matrices, and $\sigma_0$ is the identity matrix.

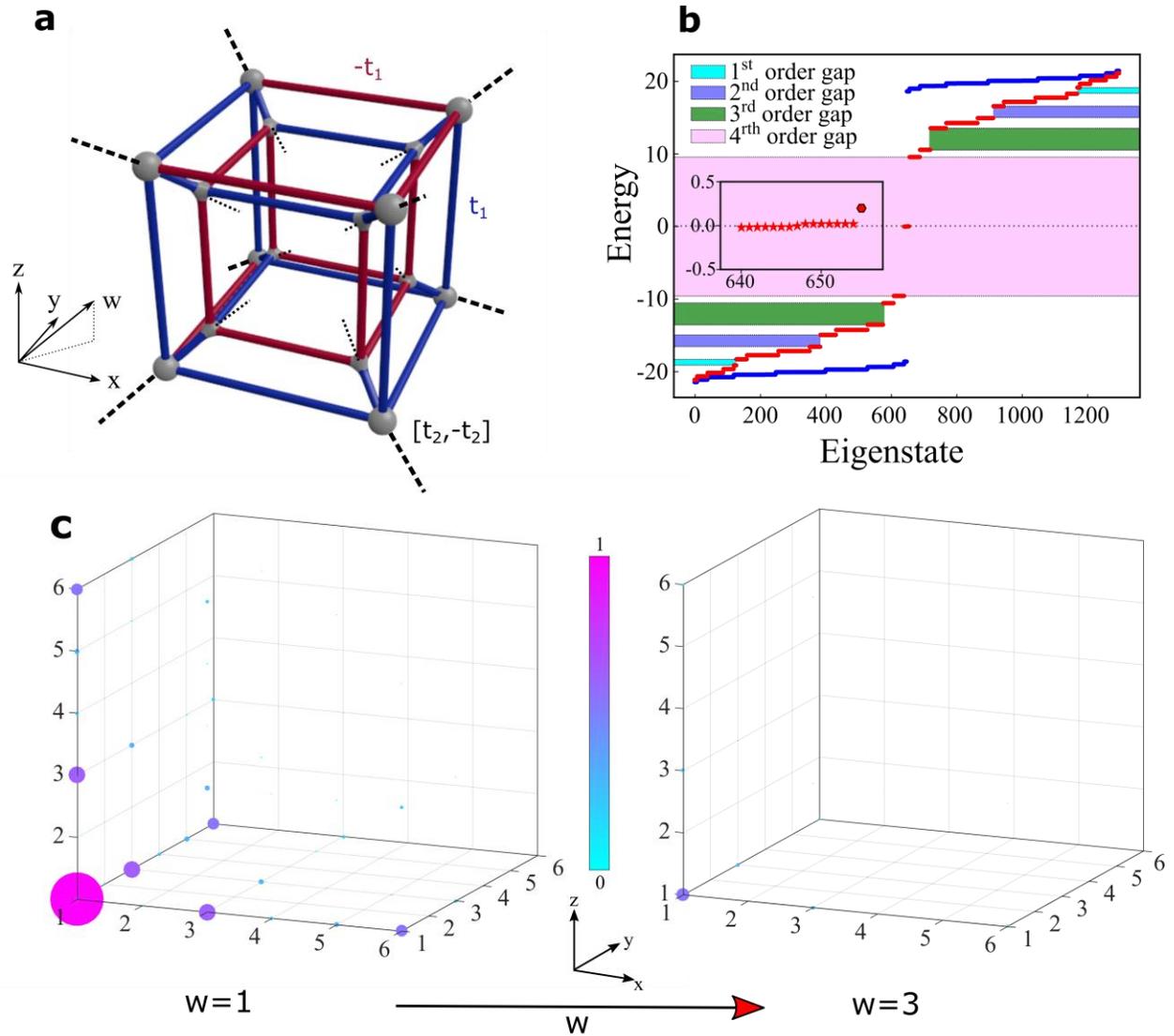

**Figure 1. Unit cell, band spectra and corner modes of 4D h-HOTI.** (a) Schematic of the unit cell of a 4D higher order topological insulator. The blue lines indicate positive hopping (zero phase) and the red lines indicate negative hopping ($\pi$ phase), so that each hypersurface has net flux of $\pi$. The black dashed lines indicate inter-cell connections to the lattice (with phases of 0 and $\pi$). (b) Band spectra for the trivial (blue dots) hyperlattice corresponding to intracell hopping amplitude $t_1$ larger than intercell hopping $t_2$ ($\frac{t_1}{t_2} = 10$). Band spectra for topologically nontrivial (red dots) higher-order phase with $t_1 < t_2$ ($\frac{t_2}{t_1} = 10$), $\delta = 0.2 t_1$. The inset figure shows a zoom-in confirming the presence of 16 corner modes. Colored regions indicate topological bandgaps of different order. (c) Amplitude distribution of corner mode across the lattice, showing the field decay and localization in all four dimensions. Each subplot shows the field distribution projected on the *xyz*-coordinate frame for specific values of $w = 1$ and $w = 3$ corresponding to the same sublattice. The corner mode shown corresponds to the energy eigenvalue indicated by the enlarged (hexagon shaped) red point in (b) inset.

The energy spectrum found from the 4D tight-binding model (TBM) of a finite (3x3x3x3x16) h-HOTI lattice with $\pi$-flux through each 2D plaquette of the unit cell is shown in Fig. 1**b**. The blue points in Fig. 1b correspond to the energy spectrum of a topologically trivial phase, with intracell hopping amplitude exceeding the intercell hopping $t_2 < t_1$, while the red dots show the energy spectrum of a topologically non-trivial HOTI phase, with $t_2 > t_1$. In the latter case, the system of finite size in all four dimensions is expected to host sixteen $4^{th}$ order topological corner states, due to the higher-order bulk-boundary correspondence[25, 26]. (See more detailed discussion in the Supplementary section S1, S2). The amplitude distribution of one of the corner modes within the h-HOTI, obtained from TBM, is shown in Fig. 1c, which confirms that this is the $4^{th}$-order state localized to the corner of the hyperlattice in all four $(x, y, z, w)$ directions. In order to make the selected corner state better noticeable, we added a small on-site potential to the corner site ($\delta = 0.2t_1$) to split this state from the rest of the corner modes.

We now map this finite four-dimensional h-HOTI onto a one-dimensional array of coupled resonators with aperiodic hopping amplitudes using the Lanczos transformation. The Lanczos tri-diagonalization transforms a general Hermitian matrix into a tridiagonal form and, therefore, allows for an immediate mapping onto a one-dimensional tight-biding model with an only-nearest-neighbor hopping distribution characterized by an inhomogeneous (aperiodic) profile. The Lanczos transformation is unitary, which ensures to preserves spectrum and eigenstate orthogonality of the original Hamiltonian, and it can be performed with a specific site of choice (i.e., a component of the eigenvector) being fixed, known as *anchor* site. The first property guarantees that the final (tridiagonal) effective 1D Hamiltonian will still be Hermitian $\widehat{\mathcal{H}}_{1D}^{eff} = \widehat{U}_L \widehat{\mathcal{H}}_{4D} \widehat{U}_L^{-1} = \sum_i \epsilon_i C_i^+ C_i + \sum_i \tau_i C_i^+ C_{i+1} + c.c.$, where $\widehat{U}_L$ is the Lanczos transformation operator, and $\epsilon_i$ and $\tau_i$ are the effective on-site energy and hopping amplitudes in the 1D array, and that the number of degrees of freedom of the 4D Hamiltonian is preserved. The second property ensures that *any* chosen site of the 1D array will have a one-to-one correspondence with one of the sites of the original higher-dimensional lattice, implying that a component of the eigenstate at a chosen site is fully preserved. This property can be used to make sure that at least one corner site in 4D is still anchored and pinned to the edge of the 1D array, implying that the corresponding corner state confined to this site in 4D will remain localized at the physical boundary of our aperiodic 1D lattice.

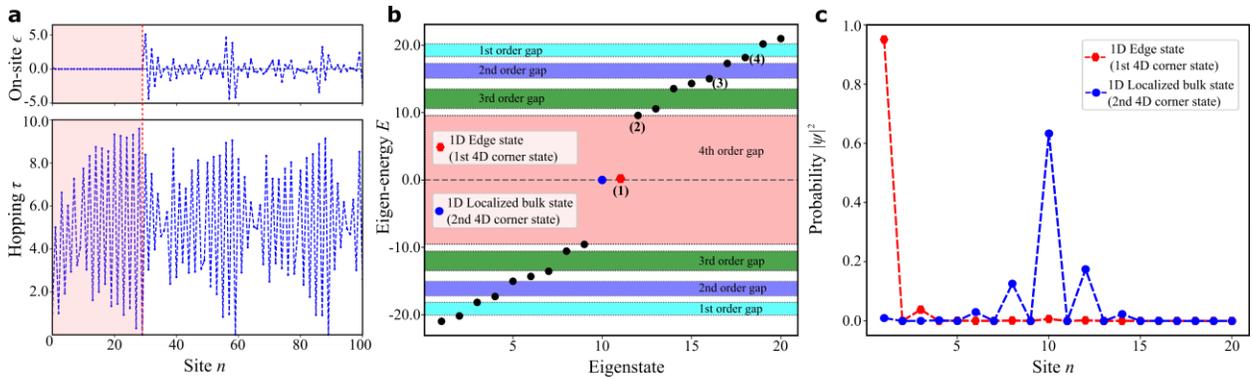

**Figure 2. Band spectra and boundary mode of the truncated one-dimensional resonator array mapped from 4D h-HOTI.** (a) Hopping amplitudes and onsite energies of the effective 1D model. (b) Band spectra of the effective 1D model truncated to 20 sites obtained using the tight-binding model. The colored regions separate modes of different order and represent topological band gaps inherited from 4D h-HOTI. (c) Amplitude distributions corresponding to the projected $4^{th}$ order topological corner mode in the 1D array, as found from tight-binding calculations.

The results of Lanczos transformation applied to the 4D Hamiltonian (Fig. 1**a**) are shown in Fig. 2**a**, where the first 100 hopping amplitudes $\tau_n$ and on-site energies $\epsilon_n$ of the effective aperiodic 1D model are plotted. The real-valued character of the 4D Hamiltonian (only phase of 0 and $\pi$ for hopping $t_n$) ensures that the Lanczos operator $\widehat{U}_L$ is an orthogonal matrix, yielding real values for onsite energies and real positive hopping amplitudes in the dimensionally reduced 1D array. Interestingly, the on-site energy $\epsilon_n$ for the first 30 sites is found to be zero, and starts to vary afterwards. Moreover, the hopping amplitude between the 30$^{th}$ and 31$^{th}$ site is 0, indicating that this part of the chain is disconnected from the rest of the array.

We numerically confirmed that the mapped 1D Hamiltonian has a spectrum identical to the one of the 4D HOTI, as expected, and that the anchored corner state appears pinned to the edge of the 1D system. Moreover, we confirmed that the inverse transformation of a 1D boundary state maps it onto corner states localized at the anchor site in 4D HOTI. It is worth mentioning that the unitary character of the transformation was enforced by implementing Lanczos biorthogonalization algorithm[37], which ensured numerical stability and one-to-one correspondence between the spectrum and eigenstates in 4D and 1D systems.

Interestingly, as the used transformation preserves only the anchor site, all other zero-energy 4D corner states appear in the bulk of the 1D lattice, but they have to remain spatially localized and topologically protected (a proof can be found in Supplement Section S3). However, since the number of higher-order boundary states (16 for the 4D case considered here) exceeds the number of boundaries of the projected 1D system (only two), most of the projected topological corner modes localize in the bulk of the 1D array. The proposed procedure thus not only offers an emulation of higher-order topological states in lower-dimensional systems, but it also unveils a new class of localized bulk resonances topologically protected by higher-dimensional symmetries of HOTIs. Indeed, the inspection of the eigenstates of the effective 1D system confirms that all other corner states of the original h-HOTI are localized modes in the bulk of the array. (See amplitude distribution of these corner modes in Supplementary Section S4).

The dimensional reduction allows for one more simplification of significant relevance for the experimental realization of $D$-dimensional HOTIs, whose number of sites increases rapidly as $(2N)^D$, where $N$ is the number of unit cells along one edge of the finite (hyper-) cubic array: the size of the system can be dramatically reduced by cutting the array at any of the weak bonds ($\tau_n \approx 0$) in the projected 1D lattice, without affecting the anchored higher-order topological corner state due to its localization at the boundary. The same holds true for the projected corner states that localize in the bulk of the 1D system, provided that the truncation is far enough from their localization site. The simulated results for the truncated 1D lattice with only 20 sites (out of 1296 original sites in 4D) are shown in Fig. 2**b,c**, and its spectrum is seen to retain a gapped character with two 4$^{th}$ order corner states pinned to zero-energy (anchored-site corner mode slightly moves away from zero-energy). Note that we intentionally cut the chain after 20$^{th}$ site to demonstrate stability of the corner modes against cutting not at exactly vanishing bond (such as after 30$^{th}$ site). Inspection of the probability distribution $|\psi|^2$ (Fig. 2**c**) confirms that one of these states is localized at the boundary (the anchor site) and the other one is localized in the bulk at the 10$^{th}$ site.

To confirm that the corner modes and their 1D projections are not affected by the truncation, we performed inverse Lanczos transformation and mapped the eigenstates of the truncated 1D system back to 4D. To this aim, we assigned zero amplitudes to the missing sites of the truncated 1D lattice to restore number of degrees of freedom. The comparison of the field distribution on the xyz hyperplane for the boundary modes of different order (indicated by numbers in Fig. 2**b**) is shown in Fig. 3, where the results for lattices truncated to 20 sites (lower panels **e-h**) and 30 sites (upper panels **a-d**) are plotted alongside. Since the case of 30 sites corresponds to the situation of 30 sites decoupled from the rest of the array, these results can be considered exact. The comparison reveals no difference in the field distributions not only for 4D corner states, but even for selected lower order hinge and surface states, and only the hypersurface states are clearly affected. Thus, the modes in general partially recover their field profiles in the projected 1d system, and appear separated by the reminiscent of bandgaps indicated by colored regions in Fig. 2**b**. This farther proves the robustness of the projected corner modes to truncations, which makes fabrication of systems supporting such states especially straightforward.

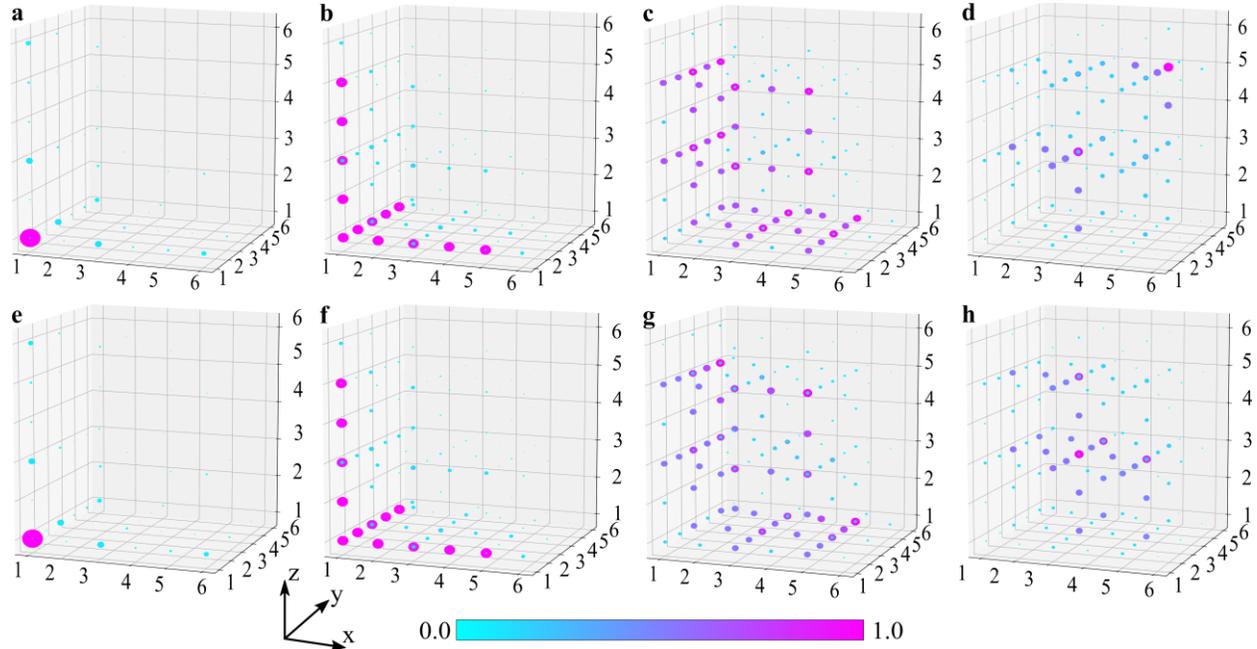

**Figure 3**. **The effect of truncation of effective 1D lattice**. **a,b,c,d** show the amplitude distributions for four different eigenstates of the 1D array of 30 sites, indicated by numbers (1 to 4) in Fig. 2**b**, projected back to 4D by the inverse Lanczos transformation. The modes clearly represent corner states, as well as reminiscent of hinge, surface, and hypersurface states in 4D, respectively. **e,f,g,h** – the same, but for the case of 1D array truncated to 20 sites. Only the projections on the $xyz$ hyperplane are shown. The distribution in $w$ direction is also found to be consistent for the two cases (not shown).

Clearly, only some of the modes are retained after the truncation of the array due to reduction in the number of degrees of freedom. As some of these states have an extended distribution in the 1D array, it is not surprising that their spectrum is slightly perturbed (due to weak coupling with the rest of the array). The rest of the states, which primarily localized in the removed part of the lattice, are dropped by the cutting procedure. We also calculated the band spectra and amplitude distribution of the effective 1D model truncated after the 30$^{th}$ site, finding that the zero energy and the corresponding amplitude distribution are not changed (See the Supplementary Section S5).

To experimentally confirm the feasibility of the dimensional reduction for the h-HOTI, we designed an acoustic aperiodic array of resonators emulating the 1D TBM using the finite-element method (FEM) software COMSOL Multiphysics (Acoustic Module). The individual resonators have the same height 3.0 cm, so that the lowest-frequency resonance, corresponding to an odd pressure profile in the vertical (axial) direction (with a single node at the center), arises at $f \sim 5720$ Hz. Coupling is introduced by adding narrow channel-waveguides connecting the resonators in the array. The coupling strength between resonators (the hopping in the TBM) is modulated aperiodically in accordance with the mapping by vertically shifting the position of the connectors with respect to the node of the resonant mode. The simulation results for the system of 20 coupled resonators are presented in Fig. 4**a**,**b**. Figure 4**a** shows the pressure field profile of the boundary mode and of the localized bulk mode stemming from the two corner states of the original 4D h-HOTI. Figure 4**b** confirms that the spectral positions of these modes are indeed localized in the mid-gap (with a small deliberate shift for the edge state), and the extended bulk modes of 1D array are gapped, which is in agreement with the results of the TBM in Fig. 2 **d**.

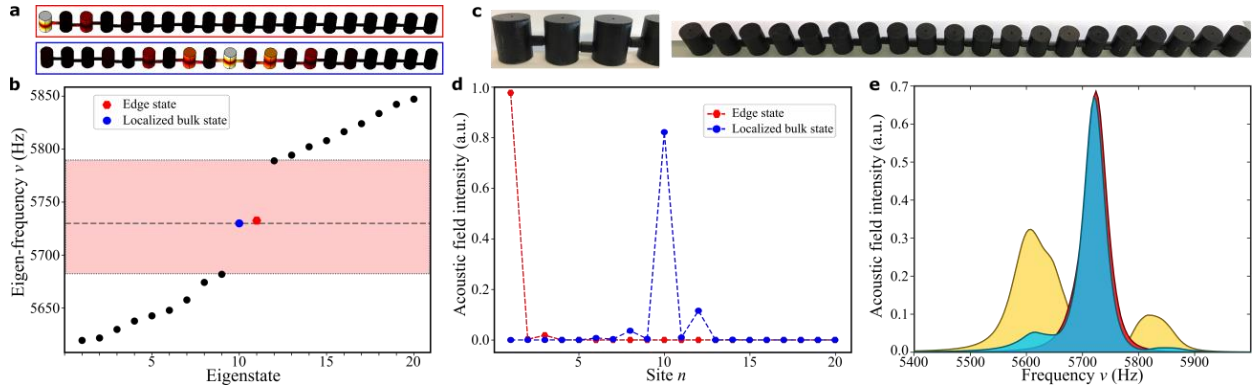

**Figure 4. Band spectra and "zero-energy" modes of the truncated one-dimensional acoustic resonator array mapped from 4D h-HOTI.** (a),(d) Amplitude distributions corresponding to the projected 4$^{th}$ order topological corner mode in 1D array, as found from first principle simulations and experimentally, respectively. (b) Band spectra of the effective 1D model truncated to 20 sites obtained using first-principle FEM calculations. (c) The effective 1D model fabricated using 3D printing with hopping amplitudes $\tau_i$'s (the coupling between the sites) varying with position. (e) Experimentally measured density of states in the array. Yellow, blue, and red regions are color-coded to represent bulk, localized bulk, and localized edge modes dominating in these regions.

The designed array of 20 coupled acoustic resonators was fabricated with the use of a high-resolution stereolithographic (STL) 3D printing (see Methods). The resonators with connectors attached to them were printed and snapped together using interlocking features deliberately introduced in the design, thus allowing for the assembly of a rigid and stable array. The assembled structure of 20 resonators is shown in Fig. 4**c**. The modes were probed with a local excitation in each resonator by placing a speaker at holes introduced on the bottom of every resonator. The strength of the local response was measured using a microphone attached to the second hole on the top of the resonators. The holes were small enough not to introduce excessive loss, but sufficiently large to probe the acoustic pressure field inside the resonators.

The frequency-response spectra for selected groups of resonators - the anchor boundary site (red band), the 10th resonator in the bulk (blue band), and averaged over all other bulk sites (yellow bands), are shown in Fig. 4**e** by color-coded bands, and clearly reveal two distinct types of states (see Methods for detailed measurement and data analysis). The one type is represented by two mid-gap states localized to the boundary and in the bulk at the 10$^{th}$ resonator, respectively; the modes that correspond to the topological corner states of the original 4D h-HOTI. Direct measurement of the field profiles corresponding to the excitation at the terminal anchor site with the frequency close to the single-resonator ("zero-energy") frequency confirms that the state is highly localized at the boundary of the array (Fig. 4d, blue line). Similarly, the second 4D corner state can be directly excited by driving the 10-th resonator at the "zero-energy" frequency, and it is found to strongly localize in the bulk. Another type of states represents delocalized states of the 1D array, which are spectrally gapped, and corresponding to non-topological hinge states of the 4D system. Despite the finite bandwidth of the resonances, of about 50Hz due to loss induced by the leakage of sound through the probe holes and absorption in the resin, the localized states projected from the 4D topological corner states clearly retain their properties, i.e., a strong localization in the array and well-defined spectral position.

The emergence of a quantized multipole moment and corner states is deeply related to the symmetries of the system, i.e., the presence of anti-commuting reflection symmetries in the 4D h-HOTI. Similarly, the "zero-energy" of the states is ensured by chiral symmetry, which stems from the fact that in the 4D h-HOTI the sites that belong to the same sublattice do not couple with each other. In addition, the Hamiltonian (1) possesses reflection symmetries that anti-commutate with each other because of the flux of $\pi$ in each plaquette of the hypercubic lattice. The chiral symmetry of the system, on the other hand, with matrix representation $\hat{\Gamma} = \sigma_3 \otimes \sigma_3 \otimes \sigma_3 \otimes \sigma_0$, is expressed as $\hat{\Gamma}\hat{\mathcal{H}}_{4D}\hat{\Gamma}^{-1} = -\hat{\mathcal{H}}_{4D}$, which ensures the overall symmetry of the spectrum and "zero-energy" of the corner states. The fact that this property is retained during the dimensional reduction implies that the chiral symmetries is preserved in new form and can be expressed as $\hat{U}_L\hat{\Gamma}\hat{U}_L^{-1}$. Indeed, due to its local character, the chiral symmetry can be written as a symmetry operator for the finite h-HOTI as $\hat{\Gamma}^{finite} = I_{N\times N} \otimes \hat{\Gamma}$, where $I_{N\times N}$ is the $N$ by $N$ identity matrix and $N$ is the number of degrees of freedom (sites), and thus is (block)-diagonal. The Lanczos transformation changes the local character of the chiral symmetry by mixing different sites (except the anchor site) $\hat{\Gamma}_{1D}^{eff} = \hat{U}_L\hat{\Gamma}^{finite}\hat{U}_L^{-1}$, thus inducing the new non-local form of chiral symmetry $\hat{\Gamma}_{1D}^{eff}\hat{\mathcal{H}}_{1D}^{eff}\hat{\Gamma}^{eff^{-1}} = -\hat{\mathcal{H}}_{1D}^{eff}$ of the effective 1D Hamiltonian. This effective chiral symmetry of the 1D Hamiltonian plays the same role as the original chiral symmetry for 4D h-HOTI and it ensures spectral stability of the modes of aperiodic 1D array. However, the non-local character of the symmetry operator $\hat{\Gamma}_{1D}^{eff}$ in 1D is reflected in the presence of correlations of parameters in different part of array, hopping amplitudes and on-site energies, within the 1D system, responsible for the "zero-energy" of the localized states. The reflection symmetries and the resultant quantized multipole moment can be similarly analyzed in the dimensionally reduced system, and the quantized multipole moment of the bulk bands in the finite array can be extracted from the respective wave functions. Thus, despite its low-dimensional character, the effective 1D system inherits the properties of higher-dimensional h-HOTI. Therefore, the projected corner states, localized either in the bulk or on the edge of the 1D array, are induced and protected by the symmetries of the original 4D system, thus ensuring their very existence and stability specific to topological systems. (See more detailed discussion in the Supplementary Section S6).

## Summary

In this work we have demonstrated a new approach to emulating higher-dimensional topological systems, such as 4D HOTI with quantized hexadecapolar multipole moment, in a dimensionally reduced 1D system. We have theoretically mapped the 4D HOTI system onto an effective 1D tight-binding model with the use of Lanczos tridiagonalization, which was then experimentally implemented in an array of coupled acoustic resonators. We have proven that higher-dimensional topological corner states arise in the effective 1D system in the form of localized resonant modes confined to the edge of the system or in the bulk. More importantly, we have shown that these modes retain their spectral features enforced by chiral symmetry and reflection symmetries, which assumes a non-local character in the 1D system, and thus establishes correlation of parameters within the array. Thus, the symmetries inherited from the original 4D system are mapped into 1D and ensure the very existence and spectral stability of the 1D projections of higher-order topological corner states. The possibility to engineer multiple localized resonances via dimensional reduction and their unique features, such as precise spectral properties and topological robustness, open remarkable opportunities for practical applications, from robust resonators to sensors and aperiodic topological lasers. More broadly, our work shows how higher-dimension topological physics may be mapped into easily realizable devices using dimensionality reduction, opening unique opportunities to implement and verify the exotic features of higher-dimension topological physics in quantum and classical photonic and phononic systems.

## Methods

*1. Structure design, 3D printing, and generic measurements* – The unit cell designs of the topological expanded lattice are shown in Fig. 2(a) with lattice constant $a_0 = 33$ mm, height $H_0 = 30.00$ mm, and radius $r_0 = 10$ mm. The connectors between the cylinders are radial channels with diameter $d_\gamma = 5$ mm. The unit cells and boundary cells were fabricated using the B9Creator v1.2 3D printer. All cells were made with acrylic-based light-activated resin, a type of plastic that hardens when exposed to UV light. Each cell was printed with a sufficient thickness to ensure a hard wall boundary condition and narrow probe channels were intentionally introduced on top and bottom of each cylinders to excite and measure local pressure amplitude at each site. The diameter of each port is 2.50 mm with a height of 2.20 mm. When not in use, the probe channels were sealed with plumber's putty. Each unit cell and boundary cell were printed one at a time and the models were designed specifically to interlock tightly with each other. The expanded structure shown in Fig. 2(a) contains 20 unit cells. For all measurements, a frequency generator and FFT spectrum analyzer scripted in LabVIEW were used.

*2. Numerical method* – Finite element solver COMSOL Multiphysics 5.2a with the Acoustic module was used to perform full-wave simulation. In the acoustic propagation wave equation, the speed of sound was set as 343.2 m/s and density of air as 1.225 kg/m^3. Other dimensional parameters of the structure are the same as the fabricated parameters.

**Data availability**

Data that are not already included in the paper and/or in the Supplementary Information are available on request from the authors.

**Acknowledgements**

The work was supported by the National Science Foundation with grants No. DMR-1809915, EFRI-1641069, and by the Defense Advanced Research Project Agency.

**Author contributions**

All authors contributed extensively to the work presented in this paper.

**Author Information**

The authors declare no competing interests. Correspondence and requests for materials should be addressed to Alexander B. Khanikaev.